\documentclass[10pt,a4paper,twocolumn]{article}
\usepackage[T1]{fontenc}
\usepackage[utf8]{inputenc}
\usepackage{lmodern}
\usepackage[margin=1in]{geometry}
\usepackage{microtype}
\usepackage{graphicx}
\usepackage{amsmath,amssymb}
\usepackage{booktabs,array,multirow,threeparttable}
\usepackage{xspace}
\usepackage{placeins}
\usepackage{adjustbox}
\usepackage[authoryear,round]{natbib}
\usepackage{xurl}
\usepackage[hidelinks]{hyperref}
\DeclareRobustCommand{\proj}{DexAttention\xspace}
\DeclareRobustCommand{\projs}{Dex\xspace}

\newcommand{\repositoryline}{\textbf{Code: }\url{https://github.com/CsRic/DexAttention}}

\title{\proj: Distributed Sparse Attention through Rapid Online Planning and Persistent GPU Execution}
\IfFileExists{latexml.sty}{\usepackage{latexml}}{\newif\iflatexml}
\iflatexml
  \usepackage{authblk}
  \author[2]{Sirui Chen}
  \author[1]{Xuehai Qian\thanks{Corresponding author: Xuehai Qian.}}
  \affil[1]{Tsinghua University}
  \affil[2]{Purdue University}
  \newcommand{\authornote}{}
\else
\author{
  Sirui Chen$^{\ddagger}$ \qquad Xuehai Qian$^{\dagger *}$\\[0.5em]
  {\normalsize $^{\dagger}$Tsinghua University \quad $^{\ddagger}$Purdue University}
}

\newcommand{\authornote}{%
  \begingroup
  \renewcommand{\thefootnote}{\fnsymbol{footnote}}%
  \footnotetext[1]{Corresponding author.}%
  \endgroup
}
\fi

\date{}
\hypersetup{
  pdfauthor={Sirui Chen and Xuehai Qian},
  pdftitle={DexAttention: Distributed Sparse Attention through Rapid Online Planning and Persistent GPU Execution},
  pdfkeywords={Distributed Sparse Attention, Context Parallelism, Online Planning, Persistent Kernel}
}
\begin{document}
\twocolumn[\begin{@twocolumnfalse}
\maketitle
\begin{abstract}
Sparse attention reduces computation, but distributed execution
may still transfer unnecessary data and incur substantial
per-operation overhead.
Mask-aware assignment and scheduling can reduce these costs,
but replanning for dynamic masks must be cheaper than the
execution time it saves.
We present \proj{}, a distributed sparse attention system
that co-designs fast online planning with persistent execution.
Its assigner uses GPU-accelerated, topology-aware clustering
to balance computational load and reduce communication.
A lightweight greedy scheduler orders coarse communication
and computation operations, while a persistent kernel on
each GPU enforces fine-grained dependencies and dispatches
ready tasks.
This division keeps detailed synchronization out of planning
while enabling overlap without repeated attention launches.
The framework supports static and dynamic per-head masks
across multiple attention compute kernels.
On a four-node, 16-GPU cluster, \proj{} achieves a
1.39$\times$ geometric-mean end-to-end training speedup
over the fastest applicable baseline across 11 configurations.
For dynamic video attention, it achieves
2.11--2.37$\times$ speedups over Ring Attention including planning.
\end{abstract}
\begin{center}\small\repositoryline\end{center}
\vspace{1em}
\end{@twocolumnfalse}]
\authornote
\section{Introduction}
\label{sec:introduction}

Long-context language models and video diffusion models make attention
increasingly expensive: its computation grows quadratically with sequence
length, while sequence-dependent activations consume GPU
memory~\citep{vaswani2017attention,peebles2023dit}.
Context parallelism (CP) distributes this work across GPUs, exchanging
queries, keys, values, or partial outputs to compute each attention
layer~\citep{liu2023ring,jacobs2023ulysses,fang2024usp,gu2024loongtrain}.
Sparse attention reduces the work itself. Sliding windows restrict
interactions by position; shared-prefix training avoids recomputing common
histories; content-dependent selection retains relevant tokens or
blocks~\citep{xiao2024streamingllm,zhang2026arealdta,yuan2025nsa,
zhao2025infllmv2,zhang2026spargeattention2trainablesparseattention}.
These workloads change both the work to distribute and the data each GPU
needs.

Sparse computation alone does not guarantee cheaper distributed execution.
A GPU may compute only selected interactions yet receive the entire KV
sequence through a ring or all-gather. Mask-aware systems such as
UltraAttn, DCP, and MagiAttention adapt work distribution and transfers to
the mask~\citep{yang2025ultraattn,jiang2025dcp,tao2025magiattention}.
We call this process \emph{planning}: assignment chooses where computation
runs, and scheduling orders its transfers and computation.
A useful plan reuses data without overloading a GPU or network link.
Two costs make this difficult for dynamic, irregular masks.

First, planning must be inexpensive relative to the
attention execution it accelerates.
Static masks allow plans to be prepared and reused,
whereas adapting plans to content-dependent masks adds work
to each layer's execution path.
A plan that reduces communication may therefore provide
little benefit if constructing it costs more than the
execution time it saves.
This makes planning latency part of the optimization
objective, alongside computation and communication.

Second, sparse execution magnifies the cost of
decomposing work into separate operations.
Irregular dependencies split attention into small
computations and transfers, each carrying launch and
synchronization overhead.
As sparsity reduces useful computation, these repeated
costs can consume a larger share of execution time.
Combining tasks into larger operations amortizes the
overhead but delays work whose inputs are already
available, limiting communication--computation overlap.

Our key observation is that runtime dependency tracking lets a coarse
schedule drive fine-grained execution. The planner can compare data
reuse and load-balance tradeoffs without predicting every transfer's
interleaving with computation.

We present \proj{}, which couples GPU clustering with greedy scheduling
and persistent execution. The assigner groups queries by shared KV
and compares load caps across nodes and GPUs using a coarse cost model.
The scheduler routes shared inputs through gateways and orders operations.
One persistent attention kernel per GPU and pass separates a controller
from optimized compute workers, dispatching ready tasks and coordinating
aggregation across static and dynamic per-head masks.

We evaluate six training and prefill workloads on 16 RTX A5000 GPUs.
\proj{} improves training by 1.39$\times$ in geometric mean over
the fastest applicable baseline across 11 configurations, and dynamic
video attention by 2.11--2.37$\times$ over Ring including planning.
With scheduler and executor fixed, its static assignments execute
1.09--2.14$\times$ faster than the evaluated mask-aware baselines'
assignments in geometric mean. Reassignment is workload-dependent:
scattered selections can make Q redistribution cost more than the KV
traffic it saves.

Our contributions are threefold:
\begin{itemize}
\item A common planning and execution framework for static and dynamic
per-head masks, multiple attention kernels, and flexible Q/KV movement.
\item A GPU clustering assigner and greedy scheduler that trade data reuse
against load balance with millisecond-scale planning.
\item A persistent executor that handles fine-grained dependencies at
runtime, allowing coarse schedules to drive overlapped attention execution.
\end{itemize}

\section{Background and Motivation}
\label{sec:background}

\subsection{Sparse Dependencies across GPUs}
\label{sec:sparse_attention}
For one attention head, queries $Q$, keys $K$, and values $V$ produce
\begin{equation}
O=\operatorname{softmax}(QK^\top/\sqrt{d}+M)V,
\label{eq:attention}
\end{equation}
where $d$ is the head dimension and $M$ is zero for permitted interactions
and $-\infty$ otherwise. Multi-head attention (MHA) uses separate Q/K/V
heads; grouped-query attention (GQA) shares each KV head among several
query heads~\citep{ainslie2023gqa}. GQA reduces KV data volume relative to
attention computation, changing the communication--computation tradeoff.

CP partitions sequence data across GPUs. Ulysses exchanges sequence
partitions for head partitions, whereas Ring Attention keeps queries
local and circulates KV shards~\citep{jacobs2023ulysses,liu2023ring}.
Later designs combine head and sequence partitioning or exploit
network hierarchy and two-dimensional data
reuse~\citep{fang2024usp,gu2024loongtrain,elango2025attention2d,chen2025mesh}.
To describe sparse CP, we partition Q and KV into \emph{chunks}.
Each tile $(i,j)$ of the interaction matrix represents computation
between $Q_i$ and $KV_j$; the mask determines which tiles are absent,
full, or partially masked (Figure~\ref{fig:mask-aware-cp}(a)).

Three representative workloads appear at the left of
Figure~\ref{fig:overview}. Static position-based patterns, such as
sink tokens plus a sliding window, remove distant
interactions~\citep{xiao2024streamingllm}. Static shared-prefix masks arise
when training on branched responses, including GRPO rollouts; sharing
histories avoids repeated prefix computation while retaining each branch's
dependencies~\citep{shao2024deepseekmath,zhang2026arealdta,wang2025treetraining}.
Dynamic masks select tokens or blocks from current activations in
LLMs and video DiTs~\citep{yuan2025nsa,zhao2025infllmv2,lai2025flexprefill,
zhang2025vsa,zhang2026spargeattention2trainablesparseattention}.
Unlike common static patterns, these selections may differ by head.
A sparse mask reduces computation, but saves communication only if it
reduces the union of inputs required by a GPU or node.

\subsection{Assignment and Scheduling}
\label{sec:mask_aware_cp}
\begin{figure}[t]
\centering
\includegraphics[width=0.76\linewidth]{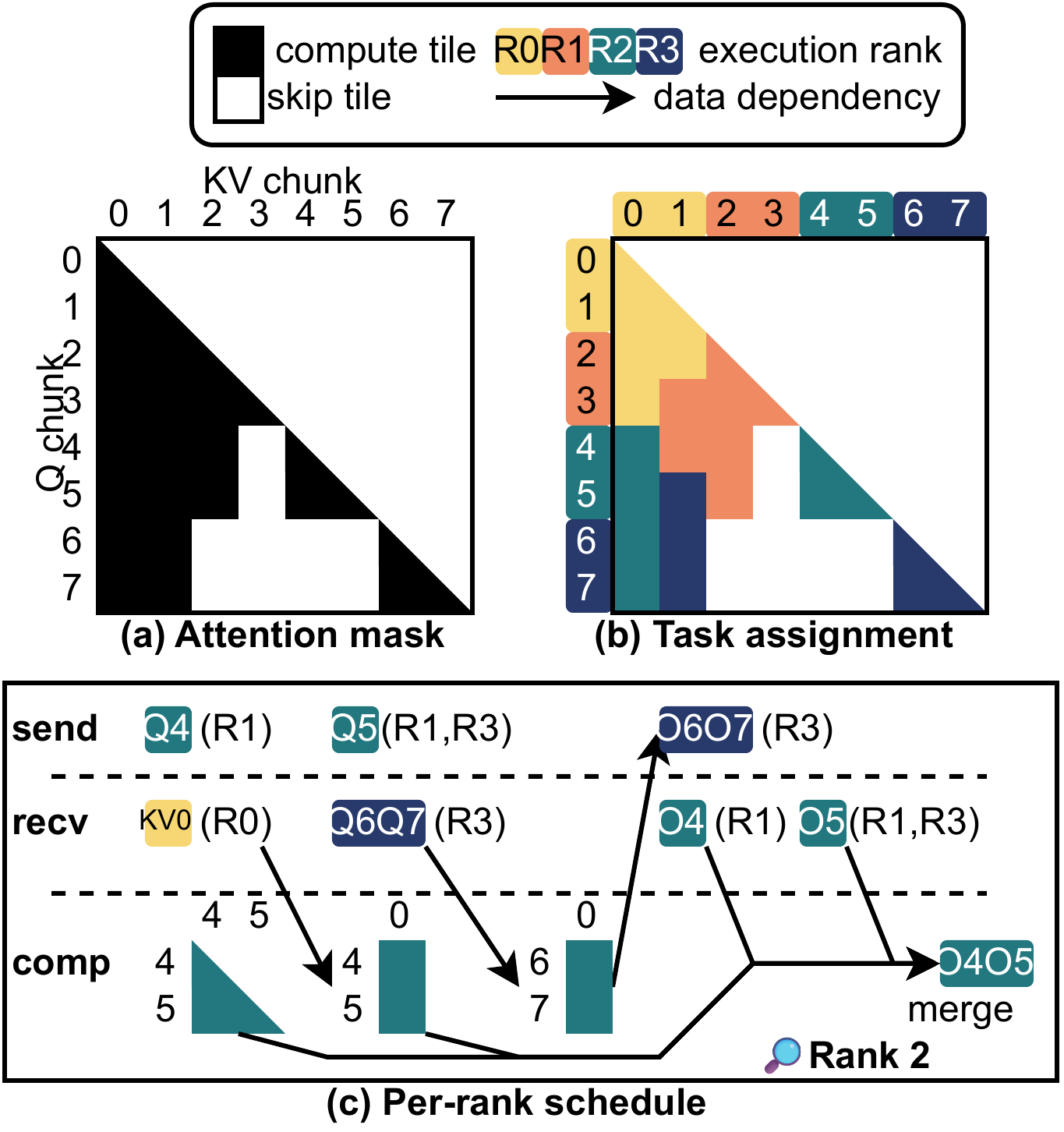}
\caption{Mask-aware CP. Axis colors show input ownership; tile colors
show execution ranks. A rank may receive either input and return partial
outputs, exposing both data reuse and scheduling dependencies.}
\label{fig:mask-aware-cp}
\end{figure}

Mask-aware CP maps these dependencies to execution through two decisions.
\emph{Assignment} maps tiles to ranks; \emph{scheduling} orders the
resulting transfers, computations, and merges.
Figure~\ref{fig:mask-aware-cp} illustrates this common abstraction.

With KV-only communication, tiles execute at their query owner.
Allowing both Q and KV movement (\emph{2D communication}) expands
assignment freedom: rank~2 in Figure~\ref{fig:mask-aware-cp}(b) computes
with $Q_{6:7}$ from rank~3 and $KV_0$ from rank~0.
Tiles along a row reuse Q, and tiles along a column reuse KV.
Colocating such work saves transfers but can concentrate computation.
The scheduler must then respect input arrival and return partial outputs
to their owners. Forward partials combine through softmax normalization
statistics~\citep{milakov2018online}; independent work can overlap.

Existing systems explore different parts of this design space
(Table~\ref{tab:system-comparison}). The evaluated UltraAttn, DCP, and
MagiAttention interfaces use shared-head masks. DCP and MagiAttention's
evaluated backends further restrict visible KV ranges, limiting fragmented
patterns. Head partitioning is a separate capability; evaluation follows these
interface restrictions (\S\ref{sec:setup}).

\begin{table}[t]
\centering\small
\setlength{\tabcolsep}{3pt}
\caption{Distributed sparse attention systems. Head par.: heads are an
assignment dimension; 2D: both Q and KV may move. Plan: measured latency
(both static passes for baselines; one dynamic pass for \projs{}).
Ranges indicate latency scale, not paired speedups. The lower three systems
are discussed but not evaluated. UA uses our adapted decomposed path.}
\label{tab:system-comparison}
\resizebox{\linewidth}{!}{%
\begin{tabular}{@{}llllll@{}}
\toprule
System & Plan & Mask scope & Head par. & 2D & Execution \\
\midrule
UA & 0.9--116 s & Shared-head & No & Yes & Decomposed \\
DCP & 6.4--111 s & Shared-head; ranges & Yes & Yes & Decomposed \\
Magi & 30--760 ms & Shared-head; ranges & No & No & Decomposed \\
\projs{} & 6--23 ms & Per-head & Yes & Yes & Persistent \\
\midrule
SparSP & --- & Video priors & No & No & Decomposed \\
FlashMask-3 & --- & Per-head; intervals & No & No & Separate kernels \\
db-SP & --- & Per-head DiT & Yes & Yes$^{*}$ & Decomposed \\
\bottomrule
\end{tabular}}
{\footnotesize ---: not measured. UA times cover successful plans.
$^{*}$db-SP redistributes Q/KV before Ulysses/Ring execution;
this does not imply arbitrary tile assignment.}
\end{table}

We discuss three additional systems but do not evaluate them.
SparSP optimizes initial placement using video-specific
profiling~\citep{sun2026sparsp}, complementary to our planning
under fixed ownership.
FlashMask-3 targets interval-structured per-head
masks~\citep{wang2026flashmask3}, while db-SP adapts
Ulysses/Ring partitioning for sparse video
attention~\citep{chen2026dbsp}.
Our comparisons focus on assignment and scheduling under
a common initial placement and on their execution costs.

\subsection{Execution and Overlap}
\label{sec:execution_motivation}
The executor realizes the plan's overlap opportunities.
In decomposed execution, coarse operations delay ready work, while fine operations
repeat launches and synchronization and may compete for GPU
resources~\citep{jeaugey2025nccl228,uccl2025transfer}.
DeepEP and COMET explore GPU-driven communication and fused
coordination~\citep{deepep,zhang2025comet}. mKernel fuses dense Ring
Attention through multiple kernel stages~\citep{mkernel}.
Persistent execution offers a basis for coordinating irregular sparse
tasks without repeated attention launches.

\subsection{Motivation}
\label{sec:motivation}
\label{sec:baseline-limitations}
Static masks allow planning to be amortized; dynamic masks require
mask-dependent decisions during execution.
Table~\ref{tab:system-comparison} shows that UA and DCP take seconds to
minutes, while Magi takes tens to hundreds of milliseconds.
With attention calls lasting tens to hundreds of milliseconds in our
experiments, these costs can consume or exceed the savings from replanning.
Their shared-head interfaces also exclude our per-head dynamic workloads.
Our focus is flexible tile assignment beyond profiled placement or
adaptation within Ulysses/Ring: can its generation cost remain small
enough to benefit each dynamic attention call?

Effective overlap creates room for simpler planning decisions:
moderate compute imbalance can be hidden behind communication, and
additional transfers can be hidden behind computation.
Assignment can therefore prioritize the dominant cost, while scheduling
leaves fine-grained readiness to the runtime. This motivates inexpensive
assignment and coarse scheduling backed by persistent execution. 

\section{Design and Implementation}
\label{sec:design}
\subsection{System Overview}
\label{sec:overview}

\begin{figure*}[t]
    \centering
    \includegraphics[width=0.96\linewidth]{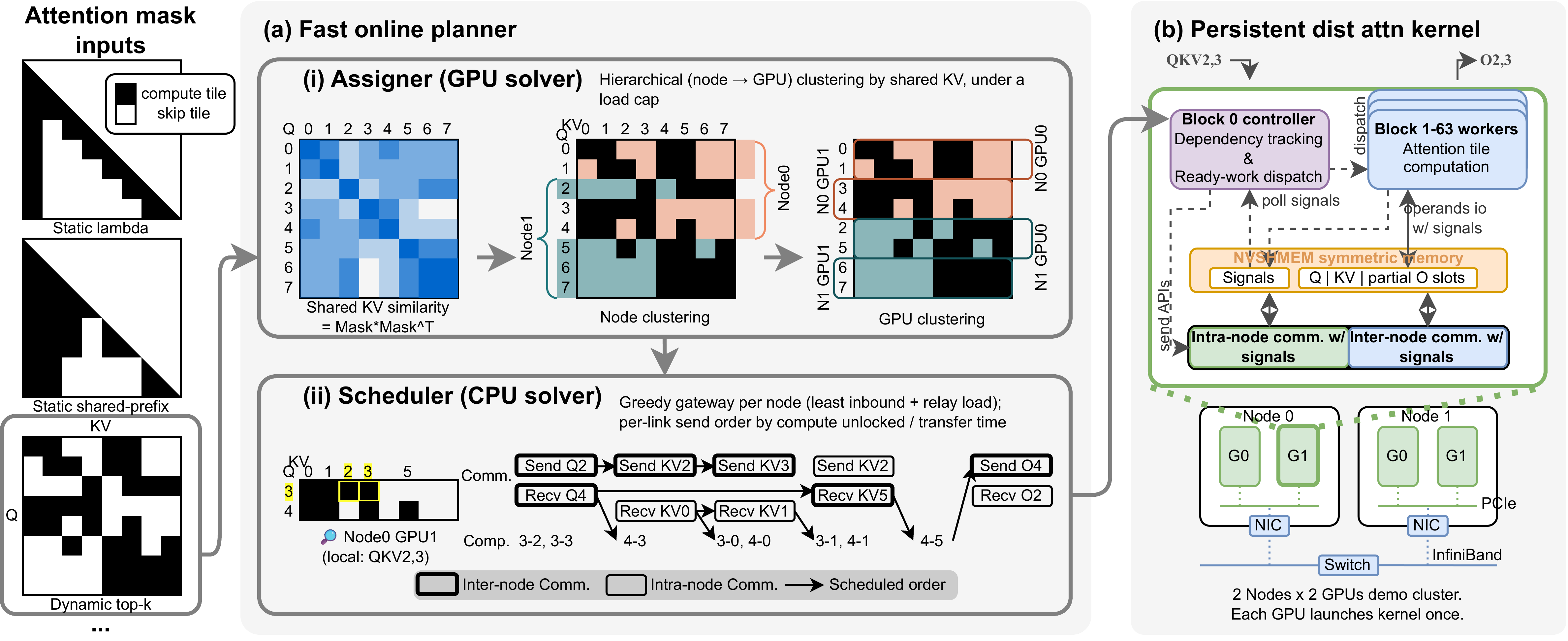}
    \caption{\proj{} overview.
    (a) GPU-accelerated assignment and CPU-based scheduling
    produce per-rank execution tables.
    (b) A persistent executor coordinates communication
    and attention computation through dependency signals.}
    \label{fig:overview}
\end{figure*}

\proj{} combines fast online planning with persistent
execution (Figure~\ref{fig:overview}).
We keep input ownership fixed under contiguous or zigzag
placement~\citep{zhuzilin2024zigzag}.
Mask-specific placement requires prior knowledge or
redistribution; instead, our planner assigns computation
under the existing ownership and includes any required
data movement in the plan.

The GPU-accelerated \emph{assigner} produces tile assignments.
The CPU-based \emph{scheduler} converts them into per-rank transfer
and compute tables. A persistent \emph{executor} consumes these tables
and dispatches tasks as their inputs become available, using one
attention kernel launch per GPU and pass.

\subsection{Fast Online Planning}
\label{sec:planner}

The planner compares batched assignments by estimated completion time,
then orders their operations without simulating execution timelines.

\subsubsection{Scope and Cost Model}
\label{sec:cost_model}
Heads with a common mask can be planned together, accounting for their
combined work and bytes. Distinct masks retain separate dependencies,
conceptually forming diagonal submatrices
(Figure~\ref{fig:clustering-assigner}(a)). The implementation batches
these submatrices by head group rather than materializing the larger
matrix. Planning chunks aggregate compute tiles; causal and other partial
tiles contribute their actual work.

An assignment determines each GPU's compute load and the
data transferred across links.
We rank candidates by their estimated completion time:
\begin{equation}
\widehat T
= \max_g \left[
    \max\!\left(T_g^{\mathrm{comp}}, T_g^{\mathrm{comm}}\right)
    + T_g^{\mathrm{tail}}
  \right].
\label{eq:cost}
\end{equation}
Compute time follows from assigned work and measured
kernel throughput; communication time follows from transfer
volume and link bandwidth, accounting for shared node NICs
and intra-node links.
Remote data shared by GPUs within a node is counted once
at its NIC, with local relaying accounted for separately.
The tail estimates computation delayed by remote inputs
and final completion overhead.
Calibration parameters are fixed across workloads.
This coarse estimate selects among candidates without
predicting their detailed execution timelines.

\subsubsection{Clustering-Based Assignment}
\label{sec:clustering}
Query chunks with overlapping KV dependencies can reuse transferred
data when computed together. The assigner merges such rows while
limiting the work concentrated at any destination.

\begin{figure*}[t]
\centering
\includegraphics[width=0.96\linewidth]{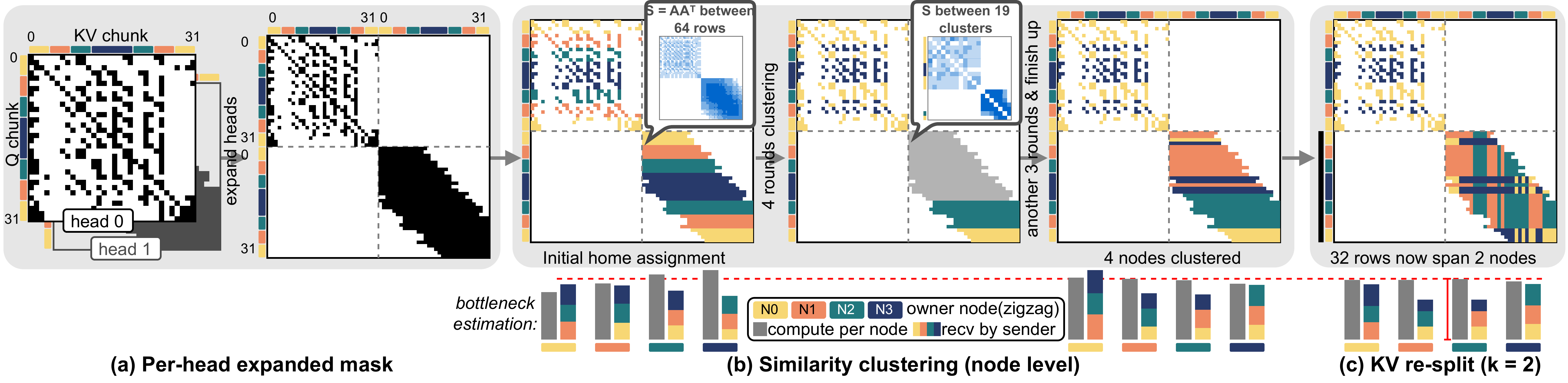}
\caption{Node-level assignment: (a) per-head masks, (b) row clustering,
(c) KV re-splitting. Axes show ownership; tile colors show execution nodes.
Gray/colored bars estimate compute/incoming-transfer time.
The GPU level repeats this procedure within each node.}
\label{fig:clustering-assigner}
\end{figure*}

\paragraph{Similarity and affinity.}
For a binary chunk mask $A$, the matrix product
\begin{equation}
S=AA^\top
\end{equation}
counts shared KV chunks, measuring reuse for equal-size chunks.
Rows start as individual clusters; mutually preferred pairs merge in
parallel, taking the union of KV dependencies and the sum of compute
work. Similarity is recomputed after each round. Figure~\ref{fig:clustering-assigner}(b) reduces 64 rows
to 19 clusters in four rounds.

Merge benefit and placement \emph{affinity} both measure saved bytes.
Merging saves the volume of shared KV chunks; affinity measures reuse
of inputs already owned or required at a destination. Placement at the
query owner also avoids Q and partial-output transfers.
Capacity constraints are checked before comparing candidates;
mutual-preference matching considers only feasible merges. A cluster is placed
when its best affinity exceeds its best merge benefit and the destination
has capacity. Otherwise it remains available for merging.

\paragraph{Reuse versus balance.}
For total work $W$ and destinations $U$, both cluster and destination
loads are capped at
\begin{equation}
C_\varepsilon=(1+\varepsilon)W/|U|.
\end{equation}
Thus, a larger $\varepsilon$ permits more reuse at the cost of greater
imbalance. Rounds stop at convergence or the eight-round limit; a final
fill places remaining clusters on lightly loaded destinations, relaxing
the cap if necessary. The cap guides clustering rather than guaranteeing
a final imbalance bound.

We batch node-level candidates with $\varepsilon\in\{0.05,0.1,0.25,0.5\}$
and a \emph{home} candidate that retains query ownership. Each node
candidate is refined across its GPUs with
$\varepsilon\in\{0.1,0.25,0.5\}$ and home. The lowest estimated completion
time selects the final assignment. This hierarchy favors reuse over the
slower shared NIC before balancing local work and transfers.

\paragraph{KV re-splitting.}
Row clustering can still replicate KV across destinations. For
communication-dominated candidates, the assigner combines row groups on
disjoint destinations and partitions their work vertically by KV
(Figure~\ref{fig:clustering-assigner}(c)). Each query may then span several
ranks, exchanging Q and partial outputs to avoid repeated KV transfers, thus enabling 2D communication.
Pairs are scored by KV bytes saved minus Q and partial-output bytes added;
non-conflicting fusions proceed in parallel. KV partitions balance work
and retain KV at its owner where possible. We try groups of at most
$k=2,3,4$ ranks, retaining a refinement round only if estimated time drops
by at least 1\%, and stop when a level gives no improvement.

\subsubsection{Lightweight Scheduling}
\label{sec:scheduler}

The scheduler converts the assignment into messages and per-GPU
compute orders. It distributes gateway work across GPUs and
prioritizes computation whose results must return to other ranks,
leaving fine-grained readiness to the executor.

\begin{figure*}[t]
\centering
\includegraphics[width=\linewidth]{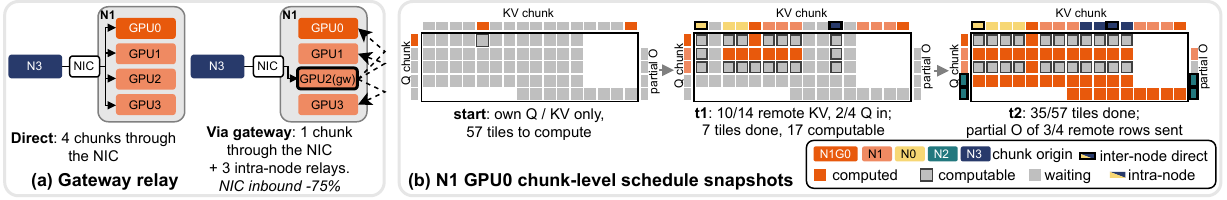}
\caption{Scheduling the assignment from
Figure~\ref{fig:clustering-assigner} with four GPUs per node.
(a) Gateway relaying avoids repeated inter-node transfers.
(b) Snapshots of N1G0 show inputs arriving through distributed
gateways and partial outputs returning before all work completes.}
\label{fig:scheduler}
\end{figure*}

\paragraph{Distributing gateway work.}
When several GPUs within a node need the same remote chunk, one
reader acts as a \emph{gateway}, receiving it once and forwarding it
locally. The scheduler selects the reader with the least combined
incoming and assigned forwarding load, distributing reception and
relaying across GPUs.
In Figure~\ref{fig:scheduler}(a), one incoming copy and three
local relays replace four copies through the NIC.
In Figure~\ref{fig:scheduler}(b), N1G0 directly receives roughly
one quarter of its required KV chunks from each of N0 and N3;
the rest arrive through gateways on other GPUs in its node.
This realizes the node-level reuse assumed by the assigner.

\paragraph{Ordering transfers.}
The scheduler divides each task's work evenly among its remote
inputs and scores each message by
\begin{equation}
\mathrm{priority}(m)=
\frac{\text{compute work attributed to }m}
     {\text{estimated transfer time of }m}.
\end{equation}
Messages on each link follow decreasing priority, while independent
links may operate concurrently.
Cumulative transfer times estimate input readiness, including
the wait for each relay's inbound message.

\paragraph{Returning partial outputs early.}
The scheduler groups computation with identical input dependencies
and orders groups by output destination: remote node first,
remote GPU next, then local.
Within each tier, groups with earlier expected input readiness
come first.
Producing remote partial outputs early allows their return
transfers and owner-side merging to overlap remaining computation.
At $t_2$ in Figure~\ref{fig:scheduler}(b), only 35 of 57 tiles
have completed, but results for three of four remote query rows
have already returned.

Sorting and linear passes determine these coarse orders;
the executor handles piece-level readiness and interleaving.

\paragraph{Planning implementation.}

\label{sec:planning_implementation}
Similarity and affinity use dense matrix products; merging, placement,
and re-splitting use batched tensor operations. CUDA graphs capture
compiled rounds, with host convergence checks. The scheduler runs on
the CPU and uploads the resulting tables. For dynamic masks, ranks first
exchange selection metadata. For W5, rank~0 additionally broadcasts
its plan to ensure consistent decisions across ranks despite
nondeterministic GPU reductions.
These exchanges are timed (\S\ref{sec:planning_results}).
\subsection{Persistent Kernel Execution}
\label{sec:executor}

The executor makes small tasks practical without requiring a detailed
static schedule. Table~\ref{tab:motivation} shows the granularity problem:
a standalone $128\times128$ attention call takes 90\,$\mu$s, while
large calls reach much higher throughput. \proj{} keeps workers resident
for an entire pass and dispatches ready pieces to them. A dense 8K-token
pass using the same tile size achieves 76\,TFLOP/s, amortizing launch and
setup rather than accelerating the underlying attention arithmetic.

\begin{table}[t]
\centering\small
\caption{Granularity costs on our cluster (16 heads, dimension 128).
Standalone FA2 and NCCL times include launch and synchronization.
Inter-node P2P uses four concurrent senders per node.
\projs{} reports amortized tile time within a dense 8K-token pass;
its communication entries are fixed overheads, excluding payload time.}
\label{tab:motivation}
\setlength{\tabcolsep}{3pt}
\resizebox{\linewidth}{!}{%
\begin{tabular}{@{}lrrr@{}}
\toprule
Operation & FA2 / compute & Intra-node & Inter-node \\
 & ms (TFLOP/s) & ms & ms \\
\midrule
$B=128$, KV 1 MiB & 0.09 (1.4) & 0.14 & 0.63 \\
$B=2048$, KV 16 MiB & 0.45 (76.8) & 0.85 & 9.30 \\
$B=8192$, KV 64 MiB & 6.54 (84.0) & 3.14 & 36.7 \\
NCCL floor ($\leq64$ KiB) & --- & 0.08 & 0.09--0.14 \\
\projs{} ($128^2$ tile) & 0.0018 (76) & 0.005 & 0.010 \\
\bottomrule
\end{tabular}}
\end{table}

\paragraph{Separate control from attention.}
Distributed control can increase register pressure in an attention
routine. Code comments in the examined mKernel Ring implementation report
spilling when stages are fused, motivating separate stage
launches~\citep{mkernel}. \proj{} instead reserves one thread block for
control (Figure~\ref{fig:overview}(b)). We launch as many blocks as SMs:
block~0 manages transfers, dependencies, and dispatch; the remaining
63 blocks execute queued attention tasks and report completion.
This keeps distributed coordination out of the attention inner loop. We integrate the workloads' optimized attention routines rather
than reimplementing their arithmetic~\citep{dao2023flashattention2fasterattentionbetter, zhao2025infllmv2}.

\begin{figure}[t]
\centering
\includegraphics[width=\linewidth]{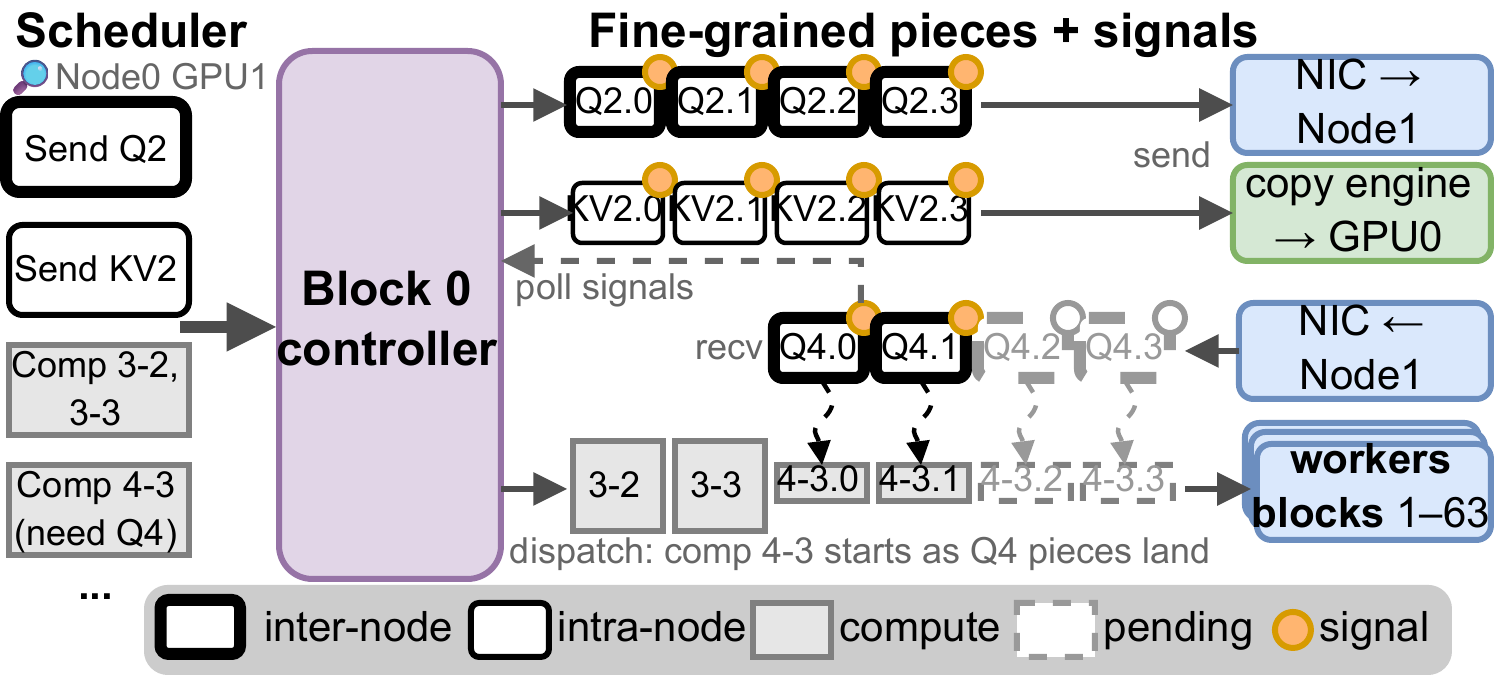}
\caption{From coarse priorities to ready tasks. The N0G1 example from
Figure~\ref{fig:overview}(a-ii) expands into piece-level signals and dispatch;
workers need not wait for a whole scheduled transfer.}
\label{fig:kernel-details}
\end{figure}

\paragraph{Interpret priorities through readiness.}
Figure~\ref{fig:kernel-details} expands the N0G1 example in the overview.
Input and partial-result slots have completion signals tagged with the
current call's epoch. The controller polls them and releases tasks whose
inputs are available. Computation 4--3 needs remote Q4: pieces Q4.0 and
Q4.1 can enable tasks while Q4.2 and Q4.3 remain in flight. Gateway
forwarding proceeds piece by piece as well. The schedule supplies coarse
priorities; signals determine the actual fine-grained interleaving.

Completed partial outputs return to their query owners and merge with
softmax normalization statistics. Backward partial gradients are summed,
including contributions to shared KV heads. The runtime tracks the
required contributions before marking an output complete. A pairwise
credit protocol prevents the next call from overwriting buffers still
used by the previous call. 

\paragraph{Coordinate two transport paths.}
For inter-node transfers, the controller issues NVSHMEM
put-with-signal operations~\citep{nvshmem}; the NIC moves data between
GPU memories. Intra-node copies use copy engines. The host pre-enqueues
them with stream waits on readiness signals, so they require no host-side
readiness polling. Both paths publish signals consumed by the controller.
Thus, one persistent attention kernel coordinates a pass, but local
copies remain separate copy-engine operations.

The kernel launch and setup take 0.07\,ms per call. Transfers avoid per-message
launches, with overheads listed in Table~\ref{tab:motivation} and
bandwidth comparable to NCCL. Among ready tasks, head-major forward dispatch encourages KV reuse
in L2 across query rows. Backward dispatch groups tasks sharing query and
output-gradient ranges. These dispatch orders improve locality among concurrently ready tasks.

\section{Evaluation}
\label{sec:evaluation}

We first evaluate end-to-end and attention performance, then examine
planning cost and assignment characteristics. Controlled substitutions
separate planning quality from execution efficiency, followed by
planning scalability and memory measurements.

\subsection{Experimental Setup}
\label{sec:setup}
\paragraph{Workloads.}
Table~\ref{tab:workloads} covers three static and three dynamic workloads.
W1 tests fixed sink-and-window sparsity; W2/W3 test shared-prefix rollouts
under MHA/GQA; W4/W6 test content-dependent LLM sparsity, and W5 tests
video DiT sparsity. W6 evaluates prefill only.

W3 comprises 24 Qwen3-4B rollout groups on LongReason, each with a prompt
and 16 responses~\citep{qwen3,ling2025longreasonsyntheticlongcontextreasoning}.
Eight groups per prompt-length split (32K/64K/128K) span sharing fractions
of 0.3--0.9; W2 reuses these masks with Qwen1.5-1.8B~\citep{qwen15}.
W4 uses InfLLM-v2 on MiniCPM4.1-8B with LongBench~v2
inputs~\citep{zhao2025infllmv2,minicpm4,bai2025longbenchv2deeperunderstanding}.
W5 uses SpargeAttention2 on VBench-prompted Wan2.1-1.3B generations
(top-$k$ 3\% union top-$p$ 20\% mass)~\citep{zhang2026spargeattention2trainablesparseattention,wan2025,huang2023vbenchcomprehensivebenchmarksuite}.
W6 uses FlexPrefill ($\gamma=0.95$, $\tau=0.1$) on Llama-3.1-8B-Instruct
with RULER/InfiniteBench inputs~\citep{lai2025flexprefill,grattafiori2024llama3,hsieh2024ruler,zhang2024infinitebench}.

\begin{table}[!t]
\centering
\small
\setlength{\tabcolsep}{3pt}
\caption{Workloads ($H_q/H_{kv}$: query/KV heads; W1--W3 static,
W4--W6 dynamic masks; W6 forward only).}
\label{tab:workloads}
\resizebox{\linewidth}{!}{%
\begin{tabular}{@{}llll@{}}
\toprule
 & Mask & Model ($H_q/H_{kv}$) & Inputs \\
\midrule
W1 & Sink + 4K window   & Qwen1.5-1.8B (16/16)  & 4 lengths, 32K--256K \\
W2 & Shared prefix      & Qwen1.5-1.8B (16/16)  & W3's 24 groups \\
W3 & Shared prefix      & Qwen3-4B (32/8)       & 24 groups, 38K--203K \\
\midrule
W4 & InfLLM-v2 top-k    & MiniCPM4.1-8B (32/2)  & 32K--128K; 144 calls \\
W5 & SpargeAttn2 hybrid & Wan2.1-1.3B (12/12)   & 480p/720p; 42 calls \\
W6 & FlexPrefill top-$p$ & Llama-3.1-8B (32/8)  & 32K--128K; 36 calls \\
\bottomrule
\end{tabular}%
}
\end{table}

\paragraph{Baselines.}
\emph{Ring} uses USP's dense KV circulation but computes only selected
tiles~\citep{fang2024usp}. \emph{All-gather} gathers KV before sparse
computation; \emph{Sparse All-gather} fetches only requested KV through
NCCL P2P without compute overlap. The mask-aware baselines are
\emph{DCP} (KaHyPar), \emph{UltraAttn} (UA, Gurobi), and
\emph{MagiAttention} (Magi, greedy area balancing)~\citep{jiang2025dcp,yang2025ultraattn,tao2025magiattention}.
Their evaluated mask interfaces restrict them to W1--W3.
\emph{\projs{} without assigner} retains query ownership while using our
scheduler and executor.

\paragraph{Reproducibility details.}
\textit{Hardware and kernels.} Unless noted, execution uses 16 RTX A5000
GPUs (24\,GB): four per node, PCIe without NVLink, and one ConnectX-6
200\,Gb/s InfiniBand NIC per node. Separate GPU/NIC CPU root complexes
limit bidirectional RDMA to 7\,GB/s per node per direction
(1.75\,GB/s per GPU with four senders); local links reach 20\,GB/s.
Faster fabrics are not evaluated. We use CUDA~12.1, PyTorch~2.5.1,
NCCL~2.21.5, NVSHMEM~3.7.1, and FlashAttention~2.8.3: FA2 adaptations
for W1--W3/W5--W6 and InfLLM-v2's kernel for
W4~\citep{dao2023flashattention2fasterattentionbetter,zhao2025infllmv2}.
Q/K/V are 16-bit; softmax statistics and partial results are FP32;
inter-node transfers use GPUDirect RDMA. W4/W6 sample layers 0/8/16/24.
Planning uses four head groups on W1--W3, KV-head groups on W4/W6,
and individual heads on W5.

\textit{Baseline settings.} DCP and Magi use their own executors.
UA's buffer-reuse and unfinished output-merging path prevents output
validation, so modified USP executes its plans in their operation order
without fusion. DCP/UA cost models are calibrated to this hardware and
executor; each search attempt gets 120\,s, retrying larger chunks on
timeout. Magi has no hardware cost model. Its varying dispatch buffers
leave poorly reusable cached allocations (4.6\,GiB allocated versus
19.8\,GiB reserved). Training clears the cache after each layer to avoid
OOM, including the 18--22\% overhead measured at 32K--64K where uncleared
runs fit. Operator tests clear it between inputs, outside timing.

\textit{Measurement.} Operator latency spans a global barrier to the
slowest GPU's completion, using the median of 10--20 warmed iterations.
Operator tests and W6 prefill use two independent runs; per-case
differences have a median of 0.6\% and a 95th percentile of 3.3\%.
Static planning is cached; dynamic tests replay masks and include
exchange, assignment, scheduling/tables, upload, execution, and W5's
rank-0 plan broadcast, but exclude selection. End-to-end timing includes
selection, including W4's compressed-key gather and W5's pooled-key
assembly. Backward planning overlaps forward execution on a worker
thread; exposed delays remain timed. Recomputation reuses masks and
forward plans. Training uses CP with
HSDP~\citep{zhao2023pytorchfsdpexperiencesscaling}, selective recomputation,
and four distinct inputs per step. A single independent training run
measures two warmed steps on W2/W3, three on W4, and four on W5;
W6 measures nine prefills.
Systems use identical input sets.
Within each workload--length or resolution group, we compute
each baseline's geometric-mean latency ratio to \proj{}
over inputs completed by both systems (paired steps for training),
excluding unsupported, timeout, and OOM cases.
The strongest baseline is the one with the smallest
group-level ratio.

\subsection{End-to-End and Attention Performance}
\label{sec:performance}
\begin{figure*}[t]
\centering
\includegraphics[width=\linewidth]{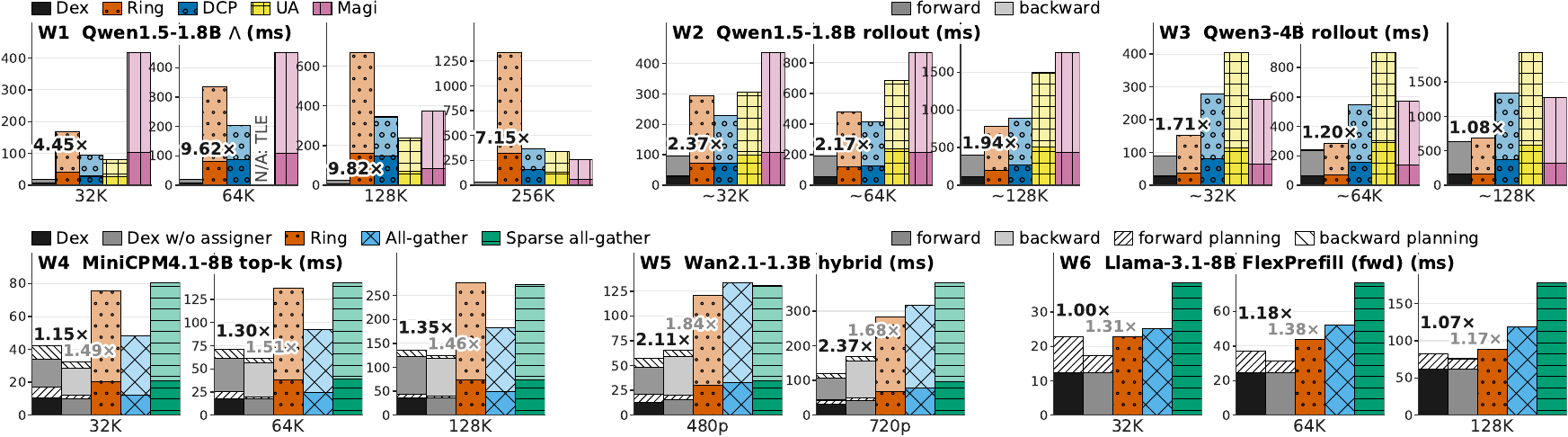}
{\small (a) Attention calls: static planning excluded; dynamic planning included.}
\includegraphics[width=\linewidth]{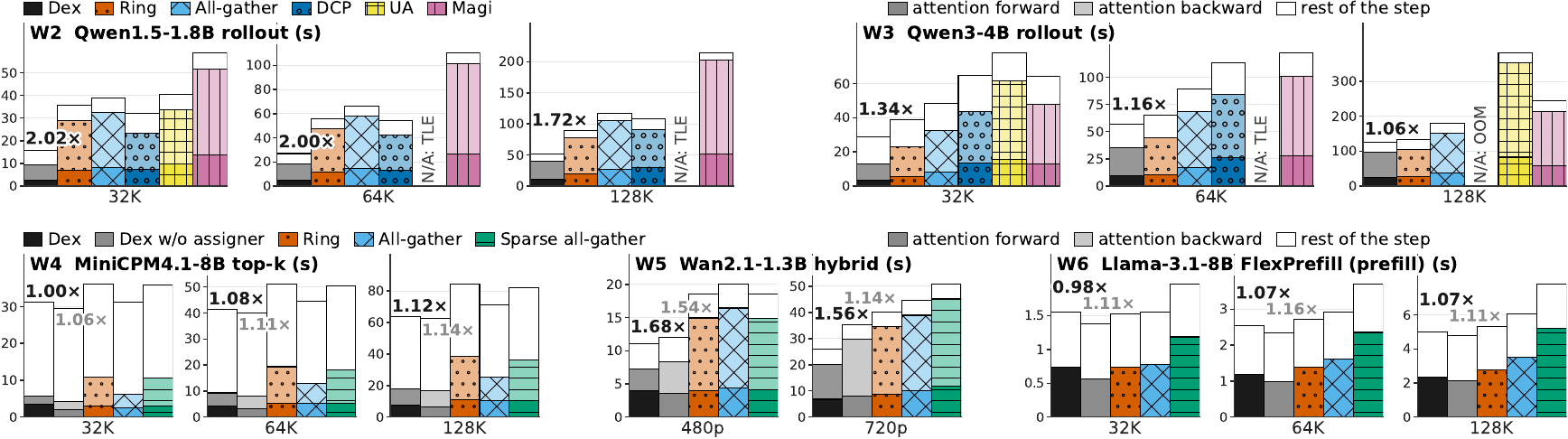}
{\small (b) Complete training steps (W2--W5) and prefill (W6).}
\caption{Performance across six workloads (geometric means).
Speedup labels compare \projs{} (black) and its variant without
assignment (gray) with the fastest applicable baseline.
Bar heights are geometric means of total latency; segments are
scaled proportionally to sum to that height. White segments show other
end-to-end work.
TLE: planning timeout; OOM: out of memory.}
\label{fig:perf}
\end{figure*}

Figure~\ref{fig:perf} compares attention latency and end-to-end performance.

\begin{table*}[!t]
\centering
\small
\setlength{\tabcolsep}{3pt}
\caption{Planning and assignment on 16 GPUs. Static time covers both
passes; dynamic time is per pass, including metadata exchange and tables.
Imbalance is maximum/mean GPU work; NIC is maximum incoming MiB per node.
Work and traffic are forward-pass geometric means on common inputs.}
\label{tab:plan}
\setlength{\tabcolsep}{3pt}
\renewcommand{\arraystretch}{0.9}
\resizebox{\linewidth}{!}{%
\begin{tabular}{@{}l
r|r
!{\vrule width 0.7pt}
rrrrrr
!{\vrule width 0.7pt}
rrrrrr@{}}
\toprule
& \multicolumn{2}{c!{\vrule width 0.7pt}}{Planning}
& \multicolumn{6}{c!{\vrule width 0.7pt}}{Imbalance}
& \multicolumn{6}{c@{}}{NIC (MiB)} \\
\cmidrule(lr){2-3}
\cmidrule(lr){4-9}
\cmidrule(lr){10-15}
System & Static & Dynamic
& W1 & W2 & W3 & W4 & W5 & W6
& W1 & W2 & W3 & W4 & W5 & W6 \\
\midrule
\projs{} & 12--19\,ms & 6--23\,ms
& 1.04 & 1.21 & 1.19 & 1.06$^{\ddagger}$ & 1.31 & 1.11$^{\ddagger}$
& 32 & 336 & 193 & 48$^{\ddagger}$ & 126 & 152$^{\ddagger}$ \\
\quad direct fetch & &
& & & & & &
& 39 & 438 & 558 & 182 & 146 & 499 \\
Ring & -- & --
& 1.03 & 1.24 & 1.23 & 1.06 & 1.05 & 1.11
& 762 & 685 & 350 & 60 & 277 & 240 \\
All-gather & -- & --
& -- & -- & -- & 1.06 & 1.05 & 1.11
& -- & -- & -- & 60 & 277 & 192 \\
Magi & 0.03--0.76\,s & --
& 1.03 & 1.01 & 1.01 & -- & -- & --
& 559 & 1355 & 699 & -- & -- & -- \\
DCP & 6.4--111\,s & --
& 1.06 & 1.44 & 1.36 & -- & -- & --
& 93 & 365 & 242 & -- & -- & -- \\
UA & 0.9--116\,s$^{*}$ & --
& 1.02 & 1.08 & 1.07 & -- & -- & --
& 183 & 1102 & 900 & -- & -- & -- \\
\bottomrule
\end{tabular}%
}

\vspace{2pt}
{\footnotesize $^{*}$Successful plans only;
14/104 UA plans time out at all tested block sizes.
$^{\ddagger}$W4, W6: \projs{} keeps every query at its owner.
\projs{} direct fetch: calculated with each GPU fetching remote inputs and returning
partial results directly to their owners, without forwarding.}
\end{table*}

\paragraph{End-to-end benefit.}
Across 11 training configurations (three lengths each for W2--W4
and two resolutions for W5), \proj{} achieves a 1.39$\times$
geometric-mean speedup over the fastest applicable baseline.
Gains vary with attention's share of the step: W4 improves by only
1.00--1.12$\times$, with non-attention work occupying 72--82\%.
On W6 prefill, \proj{} achieves 0.98--1.07$\times$ Ring's speed,
versus 1.11--1.16$\times$ without assignment. FlexPrefill's mask
selection, required by all systems, occupies 24--31\% of prefill
time with \proj{}.

\paragraph{Static masks expose reuse.}
On W1, attention is 4.45--9.82$\times$ faster than the fastest baseline.
Its local dependencies need little remote data, so dense circulation
wastes traffic and separate calls are expensive relative to useful work.
Shared-prefix masks give 1.94--2.37$\times$ on W2 and
1.08--1.71$\times$ on W3. W3's GQA already reduces KV volume relative
to computation, leaving less communication to remove.

\paragraph{Sparse computation need not imply sparse communication.}
On W5, which retains only 4--5\% of tiles, \proj{} is
2.11--2.37$\times$ faster than Ring including planning. Assignment
contributes 1.14--1.41$\times$ over the variant without it: shared
block dependencies make regrouping work worthwhile.
W4's scattered selections cover 99--100\% of preceding KV per GPU,
despite sparse computation and substantial reuse. The evaluated
assignments do not save enough KV traffic to offset Q redistribution,
so the assigner retains ownership.
\proj{} still beats All-gather by 1.15--1.35$\times$ through
relaying and overlap; skipping assignment raises this to
1.46--1.51$\times$.

W6 retains about a quarter of tiles but requires 63--71\% of
All-gather's KV bytes per GPU. Reassignment again cannot justify
moving Q. Including planning, \proj{} is 1.00--1.18$\times$ faster
than Ring. Both W4 and W6 retain ownership but pay to discover
that choice.

Sparse All-gather removes unnecessary inputs without overlap.
It matches Ring approximately on W4 but is 1.08--2.01$\times$ slower
on W5/W6: fewer bytes do not eliminate irregular NCCL P2P overhead
or hide communication.

\paragraph{Numerical agreement.}
Sampled-row checks against FP64 cover outputs and gradients on W1--W5,
including shared-KV gradient reduction. \proj{}'s worst relative L2 error
is $3.0\times10^{-4}$ on W1--W3; on W4/W5 it reaches
$4.7\times10^{-3}$, matching the corresponding All-gather result.
W6 forward outputs also match the corresponding All-gather reference. Training losses agree with
baselines within relative difference $2.7\times10^{-4}$ on W2--W4
and $2.7\times10^{-3}$ on W5.

\subsection{Planning Cost and Assignment Characteristics}

\label{sec:planning_results}

\paragraph{Online cost.}
Table~\ref{tab:plan} contrasts \proj{}'s 12--19\,ms static planning
with tens to hundreds of milliseconds for Magi and seconds to minutes
for DCP and UA. Dynamic planning takes 6--23\,ms per pass, including
metadata exchange and execution-table preparation.
Assignment takes 4.9--6.3\,ms on W5/W6; metadata exchange
and table preparation contribute additional, sometimes
comparable costs.
During W4/W5 training, backward planning runs on a worker
thread during the forward pass, leaving only forward
planning exposed: 2.6--4.9\% of a W4 step and 5--9\%
of a W5 step.
W5's hidden backward planning accounts for another
6--10\% of step time.
W6 assignment adds 5--13\% to prefill latency.

\paragraph{Balance is a means, not the objective.}
\proj{} has the lowest incoming NIC volume in Table~\ref{tab:plan}
without minimizing compute imbalance. On W5, its imbalance is 1.31
versus Ring's 1.05, but NIC traffic falls from 277 to 126\,MiB.
Magi balances work more evenly on rollouts yet incurs substantially
more NIC traffic. These metrics characterize the tradeoff;
\S\ref{sec:ablations} compares assignment quality under a common
scheduler and executor.

Relaying also contributes to the traffic reduction. For the same
assignments, measured traffic is 0.26--0.86$\times$ the direct-fetch
estimate without forwarding, with the largest reductions on W3/W4/W6.
Even without reassignment, W4/W6 receive 20\% and 21\% less traffic
than All-gather: W4 omits future KV, W6 also omits unselected KV,
and both receive shared data once per node.
Ring, All-gather, and DCP also avoid repeated inter-node transfers
through collective or two-phase communication.

\subsection{Ablation Study}
\label{sec:ablations}

We isolate planning and execution through two controls:
baseline plans on our executor, and our plans on a decomposed executor.

The static comparison uses 15 commonly supported inputs:
W1 at 32K/128K/256K (UA cannot plan 64K), and two rollouts
per length (32K/64K/128K) from W2 and W3, with valid UA
plans for both passes.
The dynamic comparison uses all 144/42/36 cases on W4/W5/W6;
W5 samples layers 0/15/29 across seven denoising steps.
Table~\ref{tab:ablation} takes geometric means within workloads,
then across workloads with equal weight.

\paragraph{Sequential component replacement.}
Table~\ref{tab:ablation} replaces executor, scheduler, then assignment;
effects are conditional on preceding replacements. Scheduler replacement
includes routing and ordering. Ring's executor and scheduler change
together because replay changes communication order. Dynamic results
reuse the main measurements: our variant without assignment retains
Ring's query ownership.

\begin{table}[t]
\centering
\small
\setlength{\tabcolsep}{3pt}
\renewcommand{\arraystretch}{0.9}
\caption{Forward latency normalized to complete \proj{}
(1.00; lower is better), geometrically averaged with equal
workload weights. Replacements are cumulative from left
to right. Static planning is excluded; dynamic planning is included.
Ring's executor and scheduler are replaced together.}
\label{tab:ablation}
\resizebox{\linewidth}{!}{%
\begin{tabular}{ccccc}
\toprule
Baseline & Native
& $\circlearrowright$ Executor
& $\circlearrowright$ Scheduler
& $\circlearrowright$ Assigner \\
\midrule
\multicolumn{5}{c}{\textit{Static masks (W1--W3)}} \\
UA   & 5.33 & 2.21 & 1.75 & 1.00 \\
DCP  & 4.25 & 1.22 & 1.09 & 1.00 \\
Magi & 4.64 & 4.00 & 2.14 & 1.00 \\
Ring & 3.55 & \multicolumn{2}{c}{1.39} & 1.00 \\
\midrule
\multicolumn{5}{c}{\textit{Dynamic masks (W4--W6)}} \\
Ring & 1.33 & \multicolumn{2}{c}{0.89} & 1.00 \\
\bottomrule
\end{tabular}%
}
\end{table}

The replacements expose different sources of overhead.
Changing the executor closes most of DCP's gap (4.25 to 1.22),
whereas Magi still takes 4.00$\times$ as long as complete \proj{};
replacing its scheduler reduces this to 2.14$\times$.

With scheduler and executor fixed, baseline assignments take
1.09--2.14$\times$ as long as \proj{}'s on static masks.
Fast clustering therefore retains competitive assignment quality,
although DCP wins on three of the 15 inputs.
The dynamic comparison reproduces the aggregate cost of assignment
identified in \S\ref{sec:performance}: normalized latency rises
from 0.89 without assignment to 1.00 with it.

\paragraph{Execution efficiency under a common plan.}
We run \proj{}'s plans on a decomposed executor across
all operator-test inputs: four W1 lengths, 24 rollout
groups each for W2/W3, and all W4--W6 cases
(Table~\ref{tab:executor-ablation}).
The control preserves assignment, node-level forwarding,
partial-result returns, and attention kernels, but uses
batched host-launched transfers, one attention call per
step, and owner-side merges.

\begin{table}[t]
\centering
\small
\setlength{\tabcolsep}{4pt}
\renewcommand{\arraystretch}{0.9}
\caption{Execution slowdown under \proj{}'s plans
(decomposed/persistent), geometric means.
Planning is excluded, subject to the W4 qualification below.
Combined-pass ratios use summed per-case latency before
geometric averaging.}
\label{tab:executor-ablation}
\resizebox{\linewidth}{!}{%
\begin{tabular}{lrrr}
\toprule
Workload & Forward & Backward & Forward + backward \\
\midrule
W1       & 2.54 & 2.56 & 2.57 \\
W2       & 2.43 & 2.38 & 2.40 \\
W3       & 2.72 & 2.10 & 2.27 \\
W4$^{*}$ & 2.34 & 3.63 & 3.18 \\
W5       & 1.58 & 1.47 & 1.50 \\
W6       & 2.47 & --   & --   \\
\bottomrule
\end{tabular}%
}

\vspace{2pt}
{\footnotesize
$^{*}$W4 ratios are upper bounds: decomposed step-list
construction remains timed, while \proj{}'s forward
deduction also removes input assembly
(0.03--0.80\,ms, at most 2\%).
}
\end{table}

With planning excluded, decomposition is
2.27--2.57$\times$ slower on W1--W3 and
1.50$\times$ slower on W5 over both passes;
W6 forward is 2.47$\times$ slower.
W4's execution-only ratios are upper bounds (table footnote);
including planning on both sides, decomposed forward is
still 1.95$\times$ slower.

\paragraph{Data reuse, relaying, and transfer order.}
On nine static inputs, assigning work by load alone
without considering shared data slows forward and
backward execution by 1.76$\times$ and 2.39$\times$.
Disabling gateway relaying slows them by
1.33$\times$ and 1.22$\times$.
Replacing transfer priorities with message-generation order
changes latency by less than 1\%.
These tests use one independent run and plans without
head splitting.
In this setting, data reuse and gateway routing matter
more than the transfer-priority heuristic.

\FloatBarrier
\begin{figure}[!t]
\centering
\includegraphics[width=\linewidth]{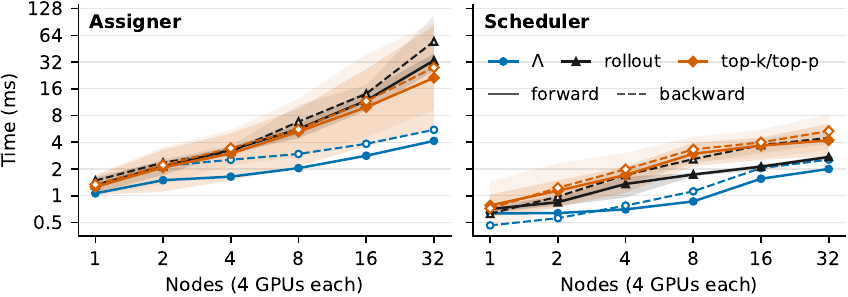}
\caption{Planning scalability: four GPUs per node, 8K tokens
per GPU, three masks per family.
Lines show geometric means; bands show sample ranges.
At $\geq16$ nodes, each rank schedules only its own node's
work; we report the maximum over sampled ranks and exclude
inter-rank metadata exchange.}
\label{fig:plan-scaling}
\end{figure}

\begin{table}[!t]
\centering
\small
\setlength{\tabcolsep}{3pt}
\renewcommand{\arraystretch}{0.6}
\caption{Estimated memory footprint (GiB, maximum across ranks): training W2--W5,
prefill W6. AG: All-gather. Unused NVSHMEM reservation is excluded.}
\label{tab:peak-memory}
\resizebox{\linewidth}{!}{%
\begin{tabular}{llrrrrrr}
\toprule
Workload & Input & \projs{} & Ring & AG & DCP & UA & Magi \\
\midrule
W2 & 32K  & 7.5 & 6.4 & 6.5 & 10.8 & 7.5 & 6.6 \\
   & 64K  & 8.9 & 7.2 & 9.0 & 13.9 & -- & 9.4 \\
   & 128K & 10.5 & 8.1 & 11.5 & 19.0 & -- & 13.5 \\
\midrule
W3 & 32K  & 11.2 & 10.2 & 10.2 & 15.5 & 11.4 & 10.1 \\
   & 64K  & 13.7 & 12.1 & 12.5 & 22.1 & -- & 12.8 \\
   & 128K & 16.4 & 14.2 & 15.5 & OOM & 17.6 & 16.5 \\
\midrule
W4 & 32K  & 13.0 & 12.2 & 11.6 & -- & -- & -- \\
   & 64K  & 15.6 & 14.5 & 13.4 & -- & -- & -- \\
   & 128K & 20.1 & 18.1 & 16.1 & -- & -- & -- \\
\midrule
W5 & 480p & 7.9 & 6.7 & 7.9 & -- & -- & -- \\
   & 720p & 9.5 & 7.7 & 10.6 & -- & -- & -- \\
\midrule
W6 & 32K  & 16.6 & 15.7 & 15.8 & -- & -- & -- \\
   & 64K  & 17.2 & 16.0 & 16.2 & -- & -- & -- \\
   & 128K & 18.8 & 17.1 & 17.1 & -- & -- & -- \\
\bottomrule
\end{tabular}%
}
\end{table}

\subsection{Planning Scalability and Memory}
\label{sec:scaling}
Figure~\ref{fig:plan-scaling} measures planning on one A5000 and its
host for up to 128 logical GPUs; execution is tested only to 16 GPUs.
At 16 nodes and above, ranks schedule their own node's work;
we report the sampled-rank maximum, excluding metadata exchange.
At 32 nodes, top-$k$/top-$p$ assignment averages 21--28\,ms,
with substantially slower cases. Reducing this variation and
distributing assignment remain future work.

Table~\ref{tab:peak-memory} reports completed runs, including groups
with incomplete performance coverage. Estimates combine the PyTorch
allocator peak with post-execution external allocations, counting
only used NVSHMEM memory.
Input, relay, and partial-result buffers add 0.8--2.4\,GiB
over Ring in training and 0.9--1.7\,GiB in W6 prefill;
all evaluated \proj{} runs fit within 24\,GB per GPU.

\section{Conclusion}
\label{sec:conclusion}
\proj{} combines GPU-accelerated assignment,
lightweight scheduling, and persistent execution for
distributed sparse attention.
Runtime dependency tracking lets coarse plans drive
fine-grained overlap.
Across 16 GPUs, it achieves a 1.39$\times$ geometric-mean
training speedup, while controlled substitutions show
that fast planning retains competitive assignment quality
and persistent execution improves performance under
fixed plans.
\FloatBarrier
\bibliographystyle{plainnat}
\bibliography{references}
\clearpage
\appendix
\section{AI Usage Disclosure}
\label{app:ai-disclosure}

We used generative AI tools, including GPT-6 family models
and Claude Opus, to assist with writing and grammar
revision, identifying and discussing related work, and
developing experimental scripts and system code,
including CUDA kernel performance tuning.
The research ideas, system design, experimental design,
and analysis were developed and conducted by the authors.
The authors reviewed the AI-assisted material and take
responsibility for the paper's content and results.
\end{document}